\documentclass[twoside]{ilcws08}
\usepackage[latin1]{inputenc}
\usepackage[dvips]{graphicx,epsfig,color}
\usepackage{wrapfig,rotating}
\usepackage{amssymb,amsmath,array}

\begin{document}
\title{
R\&D Status of FPCCD VTX} 
\author{Yasuhiro Sugimoto$^1$, Hirokazu Ikeda$^2$, 
Akiya Miyamoto$^1$, Tadashi Nagamine$^3$, \\
Yousuke Takubo$^3$,  and Hitoshi Yamamoto$^3$
\vspace{.3cm}\\
1- KEK, High Energy Accelerator Research Organization \\
Tsukuba, Ibaraki 305-0801, Japan
\vspace{.1cm}\\
2- JAXA, Japan Aerospace Exploration Agency \\
Sagamihara, Kanagawa 229-8510, Japan
\vspace{.1cm}\\
3- Department of Physics, Tohoku University \\
Sendai 980-8578, Japan\\
}

\maketitle

\begin{abstract}
As a candidate of the vertex detector for experiments at ILC,
fully depleted fine pixel CCDs (FPCCDs) are under development.
We describe the basic concept of the FPCCD and report on 
preliminary results of the performance study of the
first prototype FPCCD sensors.
\end{abstract}

\section{Introduction}
For the vertex detector in ILC experiments, pixel occupancy is
one of the most challenging issues. Due to a large amount of
electron-positron pairs created by the beam-beam interaction,
the pixel occupancy exceeds 10\% if the signal is accumulated
for one bunch train for pixel detectors with standard
($\sim 20~\mu\rm{m}$) pixel size used for the innermost layer.

In order to reduce the pixel occupancy to an acceptable
level ($< 1\%$), there are two 
different approaches studied by several groups.
One approach is to readout the signal of the sensors about 
20 times per one train. 
This method requires very fast readout speed
and very high peak power during trains. 
Another approach is to use about 20 times finer pixels
of the sensors. In this method, the signal is read out
during the train interval of $\sim 200~\rm{ms}$.
Therefore, the peak power is quite low.
The concept of the fully depleted fine pixel CCD (FPCCD) is 
an idea to realize the latter approach~\cite{sugimoto05}.

\section{Development of FPCCD sensors}
Our goal of FPCCD R\&D is to develop FPCCD sensors
with the pixel size of about $5~\mu \rm{m}$.
In addition to the small size of the pixels,
there are several challenges for the FPCCD R\&D:
\begin{enumerate}
\item[(1)] 
The sensors have to be fully depleted
to suppress the charge spread due to
diffusion.
\item[(2)]
Because very small signal ($\sim 500$ electrons) is
expected for an inclined track traversing a $5~\mu \rm{m}$
pixel, the noise level of the sensor should be
less than 50 electrons.
\item[(3)]
Readout speed should be $>10$~Mpixel/s to reduce
charge transfer inefficiency.
\item[(4)]
Power consumption (mainly by the output circuit)
should be $< 10$~mW/channel in order to
keep the detector system at $\sim -60 ~^\circ C$
with a gentle air (cool $\rm{N_2}$ gas) flow.
\item[(5)]
Horizontal register as small as the pixels
has to be put in the image area.
\item[(6)]
Wafer thickness should be as thin as $\sim 50~\mu$m
to suppress the multiple scattering.
\end{enumerate}
 
\begin{figure}
\centerline{\includegraphics[width=11cm]{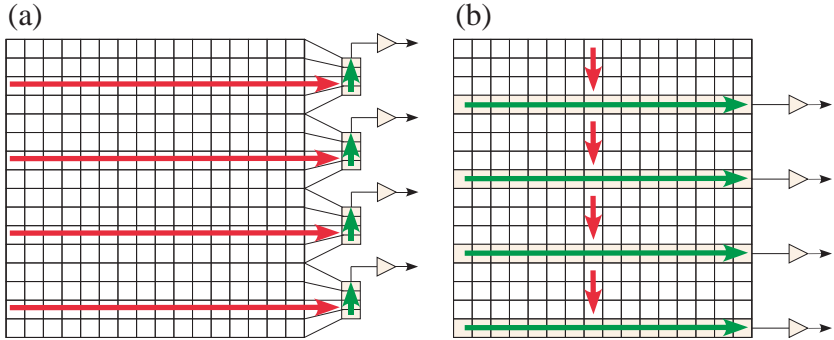}}
\caption{Two options of multi-port readout of CCDs. 
Direction of vertical (parallel) transfer is shown in red arrows
and direction of horizontal (serial) transfer is shown by green arrows.}
\label{Fig:multiport}
\end{figure}

Multi-port readout of a sensor is indispensable
to read out a large number of pixels in 200~ms.
There are two options for multi-port readout
as shown in Figure~\ref{Fig:multiport}.
In the option shown in Figure~\ref{Fig:multiport}(a),
which was adopted for the CCD sensors used 
for the vertex detector of SLD,
the number of vertical (parallel) transfer of the charge is 
larger than that of horizontal (serial) transfer.
On the other hand,  the number of horizontal 
transfer is larger than that of vertical transfer
in the option shown in Figure~\ref{Fig:multiport}(b).
Our FPCCD design is based on the option (b) from the
view point of radiation tolerance (CTE:charge transfer
efficiency).
Charge transfer inefficiency (CTI) due to traps
caused by radiation damage becomes smaller if
$1/f_{\rm{clock}}<< \tau_c$, where $\tau_c$ is
electron capture time constant ($\sim 300~\rm{ns}$
for 0.42~eV level).
For horizontal transfer clock frequency of $> 10$~MHz
and vertical transfer clock frequency of $< 1$~MHz,
the CTI associated with the vertical transfer is
larger than the CTI of horizontal transfer.
Therefore, smaller number of vertical shift realized
in the option (b) is more advantageous.
The horizontal register in option (b) has to be
put in the image area while keeping sensitivity
to charged tracks.

\begin{wrapfigure}{r}{0.42\columnwidth}
\centerline{\includegraphics[width=0.38\columnwidth]{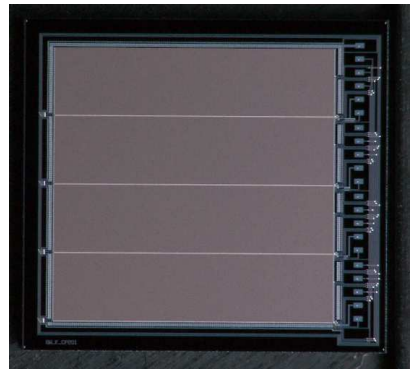}}
\caption{Photograph of a bare chip of prototype FPCCD sensor.}
\label{Fig:prototype}
\end{wrapfigure}

We have already succeeded to develop fully-depleted CCDs 
(the item (1) in the list above)
by using high resistivity epitaxial layer~\cite{sugimoto07}.
In order to tackle the challenges of (2)--(5)
among the issues listed above, we have fabricated
the first prototype FPCCD sensors collaborating
with Hamamatsu Photonics. The sensors are delivered
in packaged shape and as bare chips 
(see Figure~\ref{Fig:prototype}).
The pixel size is $12~\mu$m square in this prototype.
Each chip has $512\times 512$ pixels and is read out
through four readout ports. 
White lines seen in the image area in Figure~\ref{Fig:prototype}
correspond to horizontal registers. 

Several variants of FPCCD sensors have been manufactured.
The sensor has several different
designs of the output amplifier.
Two types of wafers with the epitaxial-layer thickness
of $24~\mu$m and $15~\mu$m are used. For the gate
oxide layer of the output transistors, two types of
chips with standard thickness and thin type are produced. 

\section{Characteristics of prototype sensors}
Basic characteristics of the prototype FPCCD sensors
are measured by the manufacturer with the operating
condition of $V_{\rm{OD}}=10~V$, $R_L=10$~k$\Omega$,
readout frequency of 10~MHz at room temperature. 
The output gain and the power consumption of the
output amplifier are 5.2--6.9~$\mu$V/electron and
10--15~mW/channel, respectively, depending on the
design of the amplifier.

\begin{wrapfigure}{r}{0.48\columnwidth}
\centerline{\includegraphics[width=0.44\columnwidth]{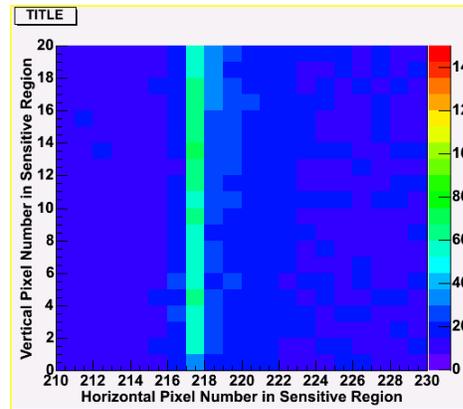}}
\caption{Image of line-focused laser light taken by a
prototype FPCCD.}\label{Fig:laser}
\end{wrapfigure}

We have tested the prototype sensors using 
line-focused laser light at KEK.
Figure~\ref{Fig:laser} shows a close-up image 
of the line-focused laser near the horizontal register 
taken by a prototype FPCCD sensor. The first (bottom) line
corresponds to the horizontal register. It can be seen
that the horizontal register is sensitive to light.
Actually, the signal of the horizontal register
is significantly smaller than standard pixels
because the horizontal register is partially covered
by aluminum layer for gate clocking.

In the prototype FPCCDs, large dark current
on both horizontal edges was observed. 
Another problem of charge injection into first few tens
of pixels of first few tens of lines was found
when vertical clock is held ``high'' for 20~ms before 
reading out a frame.
In the next batch 
of prototype in 2009, the latter problem is expected 
to be resolved. In addition, increased full-well
capacity is planned to be achieved in the next prototype.

\section*{Acknowlegements}
This work is partly supported  by  Grant-in-Aid
for Scientific Research No.~17540282 and No.~18GS0202 from
Japan Society for the Promotion of Science (JSPS),
and No.~17043010 and No.~18034007 from Ministry of
Education and Science in Japan.


\begin{footnotesize}



%

\end{footnotesize}


\end{document}